\documentclass[12pt,a4paper]{article}
\usepackage{epsfig}
\usepackage{epic}
\usepackage{float}
\usepackage{afterpage}
\restylefloat{figure}
\renewcommand{\baselinestretch}{1.7}\normalsize

\newcommand{\be}{\begin{equation}}
\newcommand{\ee}{\end{equation}}
\newcommand{\bea}{\begin{eqnarray}}
\newcommand{\eea}{\end{eqnarray}}
\newcommand{\klgl}{\:\hbox to -0.2pt{\lower2.5pt\hbox{$\sim$}\hss}
{\raise3pt\hbox{$<$}}\:}
\newcommand{\grgl}{\:\hbox to -0.2pt{\lower2.5pt\hbox{$\sim$}\hss}
{\raise3pt\hbox{$>$}}\:}
\begin{document}
%
%
\markboth{ }{ }
\renewcommand{\baselinestretch}{1}\normalsize
\vspace*{-2cm}
\hfill TUM-HEP-317/98

\hfill SFB-375-299

\vspace*{2cm}
\bigskip
\bigskip
\begin{center}
{\huge\bf{Critical behavior of $\varphi^4$-theory\bigskip\\
from the thermal renormalization group}} 
\end{center}
\bigskip
\begin{center}
{\Large{Bastian Bergerhoff}}\footnote{
bberger@physik.tu-muenchen.de}$^,$\footnote{This work was supported by 
the "Sonderforschungsbereich 375-95 f\"ur
Astro-Teilchenphysik" der Deutschen Forschungsgemeinschaft.}\vspace*{0.3cm}\\
Institut f\"ur Theoretische Physik \\
Technische Universit\"at M\"unchen \\
James-Franck-Strasse, D-85748 Garching, Germany
\end{center}
\setcounter{footnote}{0}
\bigskip
\vspace*{3cm}\begin{abstract}
\noindent
We discuss the universal critical behavior of a selfinteracting 
scalar field theory at finite temperature as obtained from 
approximate solutions to nonperturbative renormalization group (RG)
equations.
We employ a formulation of the RG-equations in real-time formalism 
which is particularly well suited for a discussion of the thermal
behavior of theories which are weakly coupled at $T=0$.
We obtain the equation of state and critical exponents of the 
theory with a few percent accuracy even for a relatively simple
approximation to the exact renormalization group equations.
\end{abstract}
\newpage
\renewcommand{\baselinestretch}{1.1}\normalsize
\noindent {\bf{1.}} Field theory at nonvanishing temperature has in recent
years attracted a lot of interest. 
It is relevant on widely separated scales, starting from the physics of the
early universe, where phase-transitions may play a role in inflation, the
formation of density inhomogeneities and the creation of the matter-antimatter
asymmetry observed today, via the by now also experimentally studied physics of
the quark-gluon plasma to critical phenomena in solid state systems.
Thermal energies involved range from $10^{15}$ GeV down to a few meV.
On the other hand, the number of open questions is still large compared to
other fields of theoretical physics.
Even for systems that can be treated as equilibrium systems, there are a
number of technical problems that have to be overcome.
This holds already for very simple models like the selfinteracting scalar
theory near its critical temperature studied in this letter.

Perturbation theory at nonvanishing temperatures is
plagued by infrared problems. 
This  is mainly due to the fact that the dynamics of the theory at high
temperatures is dominated by classical (thermal) fluctuations, which may be
described by a three-dimensional effective theory.
The infrared behavior of the three-dimensional theories prevents us from
studying the critical behavior in straightforward perturbation theory.
Indeed, simple one-loop resummed perturbation theory predicts a first order
phase-transition for the $O(N)$-symmetric scalar theory
\cite{PertFirstOrder} -- a result that is known to be wrong.

There are a number of approaches available that cure the problems of
straightforward perturbation theory.
Most of these methods use variants of the renormalization group (RG) idea.
The "environmentally friendly RG" 
(\cite{Chris1} and references therein)
is closely related to the field theoretic (Callen-Szymanzik type) RG and as
such in essence perturbative, avoiding however the above-mentioned infrared
problems.
This makes the method very well suited to study the critical behavior of
systems where the basic degrees of freedom do not change dramatically through
the phase transition, as demonstrated e.g.~by the results for the critical
exponents of scalar theories like the one studied here
\cite{Chris1}.

Another approach is the so-called "auxiliary mass method" suggested in 
\cite{Drummondetal} 
and applied to the critical behavior of scalar theories in
\cite{Satoetal}.
This method is in principle nonperturbative and thus, like the thermal
renormalization group approach employed in this letter and discussed below,
suffers from the need to make approximations that may in general only be tested
a posteriori.
Results for the critical indices of selfinteracting scalar theories are
reported in 
\cite{Satoetal} 
and are comparable to the results of the present study (see below).

The standard approach of statistical mechanics is the Wilsonian renormalization
group
\cite{Wilson}.
The basic idea of the Wilsonian RG may be implemented in
different ways.
It has been introduced in a field-theory context by Polchinski and others
\cite{Polchinskietal}
mainly to provide a physical understanding of
renormalization.
In recent years this approach has been applied to different problems and is now
also known as the "Exact renormalization group (ERG)"-approach.
It takes no recourse to perturbation theory and has been used to study strongly
interacting problems both at vanishing and at nonvanishing temperature.
Applications to critical phenomena are commonly done -- in the spirit of the
above mentioned effective theories -- in
the framework of three-dimensional theories
\cite{CN1}-\cite{Morris2},
although some work using a formulation of thermal field theory in the
"imaginary time"-formalism has also been done
\cite{ERGITF}.

Recently, an implementation of the Wilsonian renormalization group for thermal
field theory in the real-time formalism has been proposed
\cite{DAP1} (we will henceforth denote this formulation as the "thermal
renormalization group" or TRG).
Even though on a formal level the connection between the imaginary-time and the
real-time formulation of thermal field theory is well understood
\cite{TFTreview}
and all physical quantities may be computed in any of the two approaches,
if one is interested in time-dependent quantities it is often convenient to be
able to directly work in a real-time approach instead of having to do
analytical continuation of Green-functions which may be involved or even
impossible if the results are only numerical (as is typically the case in
applications of the RG).
Furthermore, in the real-time
formulation of thermal field theory there is a clear separation between
"thermal" and "quantum" fluctuations. The two-point functions may be separated
into a part identical to the propagator at vanishing temperature and a part
which contributes only at $T>0$.
In the formulation proposed in 
\cite{DAP1},
only the thermal fluctuations are treated by renormalization group methods.
The quantum fluctuations are assumed to be integrated out, i.e. one starts
from the full physical effective action of the theory at vanishing temperature.
This feature may seem to be a shortcoming, since it essentially prevents
us from studying theories which are strongly interacting already at vanishing
temperature\footnote{Although in principle the effective action of such
models could be obtained by nonperturbative methods (lattice, ERG-equations at
$T=0$,
etc), this is in practice a formidable task for realistic models like QCD.}.
On the other hand, having an exact renormalization group equation with the
physical theory as boundary condition turns out to be a major advantage if one
is interested in the thermal behavior of weakly interacting models in the
vicinity of a phase transition which is second or weakly first order and thus
not accessible in perturbation theory.
In this case, one obtains the thermal behavior directly in terms of physical
quantities (masses, couplings etc) instead of connecting it to the scheme-dependent
parameters of the action of the theory.
The usual formulation of the ERG defines its own renormalization scheme and
the connection between physical couplings at $T=0$ and the parameters of the
action has to be obtained through explicit calculation
\cite{UliSchemes}.
This typically involves ambiguities which are only fixed at the two-loop level
in a perturbative language. 
Formal two-loop exact calculations in a nonperturbative scheme like the exact
renormalization group are however notoriously difficult
\cite{Thomas2lp}.

Another advantage of the TRG is its explicit gauge-invariance
\cite{DAP2}.
In the Wilsonian renormalization group approach one introduces an external
scale into the theory which is used to separate "hard" and "soft" modes in
order to be able to consecutively integrate out "momentum shells".
This procedure violates gauge-invariance\footnote{One may work in the
background-field formulation, keeping manifest gauge-invariance with respect to
the background-gauge transformations
\cite{ChristofMartin1}.},
and one obtains generalized Slavnov-Taylor identities imposing relations
between the terms in the effective action
\cite{MSTI}.
Although from a formal point of view everything is well understood, it is
often not trivial to implement these constraints in actual calculations.
In the TRG-approach, gauge-invariance is maintained since the separation
between hard and soft modes only concerns on-shell fields, namely the thermal
fluctuations.
This feature is very advantageous for practical calculations, since it severely
restricts the possible form of solutions to the renormalization group
equations.

These points make the TRG the ideal tool to study the critical behavior of (at
vanishing temperature) weakly interacting models beyond the universal aspects,
which may be calculated in the framework of three-dimensional effective
theories. 
There are a number of interesting problems which are weakly interacting at
vanishing temperature but are nevertheless not accessible through perturbation
theory at temperatures close to the critical one.
As stated above, the simplest examples are $O(N)$-symmetric selfinteracting
scalar theories which are known to exhibit second order phase-transitions. 

Another important example is the electroweak sector of the standard model,
which is related to the observed asymmetry between matter and antimatter in the
universe
(see \cite{MishaReview} for a review).
This model is weakly interacting at zero temperature for not too large
Higgs-masses, but its critical behavior is not described by perturbation
theory (unless the Higgs-mass is very small).
Around the critical temperature, the model is effectively strongly interacting 
for a large range of Higgs-masses
\cite{EWPT1}.

Finally, the TRG may be used to study real-time quantities like thermal damping
rates 
\cite{P}
etc.
One could even hope to be able to access non-equilibrium quantities along
these lines.

A prerequisite for the study of thermal phenomena in the framework of the
TRG is the existence of useful approximation schemes
for the solution of the exact renormalization group equation.
This letter presents a study of the one-component selfinteracting scalar
theory as a testing-ground for such nonperturbative approximations.
The critical behavior of this theory has been excessively studied with the
help of the three-dimensional model. 
The theory is in the universality class of the Ising model and the critical
exponents are known to high accuracy from lattice calculations, the
$\epsilon$-expansion, ERG-studies etc
(for a review, see \cite{ZJ}).
A quantity which is more difficult to access is the critical equation of state
(EOS) of the theory.
Recently, also the EOS has been obtained from lattice studies
\cite{Tsypin}
and the ERG in three dimensions
\cite{JNC}
(in a slightly different context, the EOS has also been calculated using
environmentally friendly renormalization
\cite{Chris2}).
We aim at deriving the critical behavior not from the effective
three-dimensional model, but directly from the finite-temperature theory.
The results will show that one can give useful approximations to the exact
renormalization group equations in the case studied and will give a
nonperturbative example of dimensional reduction. 
We will not be able to improve on other methods for the critical exponents
within the approximations performed here. 
Nevertheless, our results are accurate to a few percent and do not rely on
universality, thus proving the usefulness of the TRG for the study of critical
systems and opening the way to obtain nonuniversal quantities like
critical amplitudes directly in terms of physical quantities.\vspace*{0.5cm}

\noindent {\bf{2.}} Before presenting the results, we will briefly describe the
formulation of the thermal renormalization group as proposed in 
\cite{DAP1}.
The formulation of the TRG-equations relies on the closed-time path (CTP)
approach to thermal field theory
\cite{TFTreview}.
In the operator formalism, one starts with the usual expansion of fields in
terms of creation and annihilation operators $a^\dagger_k$ and $a_k$.
Operator averages are now thermal averages, and we have 
\bea
<\!\!a^\dagger_k a_{k'}\!\!>_T &=& (2 \pi)^3 2 \omega_k N(\omega_k)
\delta(\vec{k} - \vec{k'}) \nonumber \\
<\!\!a_k a^\dagger_{k'}\!\!>_T &=& (2 \pi)^3 2 \omega_k \left[ 1 + N(\omega_k)
\right] \delta(\vec{k} - \vec{k'})
\label{averages}
\eea
where $\omega_k = \sqrt{\vec{k}^2+m^2}$ and $N(\omega_k) = \left( e^{\omega_k /
T} - 1 \right)^{-1}$.
In order to divide the thermal modes in to "hard" and "soft" ones, one 
modifies the thermal part of these averages, i.e. the part proportional to the
thermal distribution function $N(\omega_k)$ by introducing a cutoff-function 
$\tilde{\Theta}(|\vec{k}|,\Lambda)$ as
\bea
<\!\!a^\dagger_k a_{k'}\!\!>_T &=& (2 \pi)^3 2 \omega_k N(\omega_k)
\tilde{\Theta}(|\vec{k}|,\Lambda) \delta(\vec{k} - \vec{k'}) \nonumber \\
<\!\!a_k a^\dagger_{k'}\!\!>_T &=& (2 \pi)^3 2 \omega_k \left[ 1 + N(\omega_k)
\tilde{\Theta}(|\vec{k}|,\Lambda) \right] \delta(\vec{k} - \vec{k'})
\label{cutoffaverages}
\eea
where $\tilde{\Theta}(|\vec{k}|,\Lambda)$ is 1 for $|\vec{k}| \geq \Lambda$ and
rapidly vanishes for $|\vec{k}| < \Lambda$.
This has the effect of suppressing thermal modes with three-momenta small
compared to the external scale $\Lambda$ -- the external scale acts as infrared
cutoff for the thermal fluctuations.
The formulation of a generating functional for 1PI Green-functions proceeds as
usual, except that the two-point function appearing in the action now depends
on the cutoff $\Lambda$. 
We thus obtain an effective action $\Gamma_\Lambda$ which interpolates between
the physical effective action $\Gamma(T=0)$ of the zero-temperature theory as
$\Lambda \rightarrow \infty$ (no thermal fluctuations included) and the full
effective action of the theory at finite temperature $\Gamma$ as $\Lambda
\rightarrow 0$ (all thermal fluctuations integrated out).
 The explicit form of the two-point function $D_\Lambda^{-1}$ and the generating
functional $\Gamma_\Lambda$ may be found in 
\cite{DAP1}.

In the end of the calculation one will usually be interested in the limit
$\Lambda \rightarrow 0$, and in the spirit of the renormalization group one now
proceeds to deriving a differential equation for the dependence of
$\Gamma_\Lambda$ on the external scale.
Quite analogously to the exact renormalization group equation proposed in 
\cite{Christof}, 
the thermal renormalization group equation takes the form
\cite{DAP1}
\bea
\Lambda \frac{\partial}{\partial \Lambda} \Gamma_\Lambda[\Phi] &=& \frac{i}{2}
{\mathrm{Tr}} \left\{ \Lambda\frac{\partial}{\partial \Lambda} D_\Lambda^{-1}
\left( D_\Lambda^{-1} + \frac{\delta^2 \Gamma_\Lambda[\Phi]}{\delta
\Phi \delta \Phi} \right)^{-1} \right \}
\label{TRG}
\eea
(\ref{TRG}) is an exact equation and one faces the usual problem of not being able to
solve it without approximations.
In the present study, we will work in lowest order of an expansion of the
effective action in powers of derivatives.
The derivative expansion effectively may be seen as an expansion in the
anomalous dimension of the fields, and we expect it to work rather well in the
scalar theory under study here
\cite{CN1}.
We thus proceed by defining the functional $\bar{\Gamma}_\Lambda[\phi]$ through
\cite{NS} ($\Phi_1$ and $\Phi_2$ are the physical and the "thermal ghost" field
resp.)
\bea
\frac{\delta \bar{\Gamma}_\Lambda[\phi]}{\delta \phi} = \left. \frac{\delta
\Gamma_\Lambda[\Phi]}{\delta \Phi_1} \right|_{\Phi_1 = \Phi_2 = \phi}
\label{Gammabar}
\eea
For this functional we make the ansatz
\bea
\bar{\Gamma}_\Lambda[\phi] = \int d^4x \left[ \frac{1}{2} \left( \partial \phi
\right)^2 + \frac{1}{2} m^2 \phi^2 - V_\Lambda(\rho) \right]
\label{ansatz}
\eea
where (we drop the subscript in the following) $V(\rho)$ is the effective
potential and $\rho = \frac{\phi^2}{2}$. Up to a field independent term, the 
$\Lambda$-dependence of $V$ may be derived from the flow-equation (\ref{TRG})
and one finds, using for the cutoff function $\tilde{\Theta}(|\vec{k}|,\Lambda)
= \Theta(|\vec{k}|-\Lambda)$,
\cite{DAP1}
\bea
\Lambda \frac{d V(\rho)}{d \Lambda} &=& -T \frac{\Lambda^3}{2\pi^2} \ln \left\{
1 - \exp \left(-\sqrt{\frac{\Lambda^2}{T^2} + \frac{V' + 2 \rho V''}{T^2}}\right) \right\}
\Theta(\Lambda^2 + V' + 2 \rho V'')
\label{feqV}
\eea
This equation has to be solved numerically.
One way to proceed is to expand the effective potential around some field-value
$\bar{\rho}$, truncate the expansion at some finite order, and thus obtain a
finite system of coupled ODEs for the "couplings" $V^{(n)}(\bar{\rho})$
(this approach is known as the "local polynomial approximation" or LPA).
This approach was followed for the present flow-equation in 
\cite{DAP1}, 
where values for the critical exponents $\delta$, $\nu$, and $\eta$ where
obtained (see table below).
In the ERG-formulation in three dimensions, critical exponents where also
derived along these lines 
\cite{CN1}.
However, it turns out that the results obtained in the TRG-approach using the
LPA are relatively poor compared to the results obtained from lattice studies
or higher order calculations in the $\epsilon$-expansion.
This should not be connected to the derivative expansion, where we expect
errors of the order of the critical exponent $\eta$, being $\approx 0.03$ for
the one-component theory.

In this letter we will thus study the numerical solution to the full partial
differential equation (\ref{feqV}).
This has the further merit that we will be able to give the approximate
equation of state, namely the function $\frac{d V(\phi)}{d \phi} = \phi
V'(\rho)$ in the vicinity of the critical temperature.
We use the algorithm proposed in 
\cite{Matching}: 
We discretize in $\rho$ and use the flow-equations for $V'(\rho)$ and
$V''(\rho)$ at each point. 
The flow-equations involve the derivatives of $V$ up to $V^{(4)}$, 
and we obtain these derivatives from matching relations by imposing continuity
of $V'$ and $V''$ at the points.
In the (low temperature-) broken phase, it turns out convenient to also follow
the $\Lambda$-dependence of the minimum of the potential, given by
\bea
\Lambda \frac{d \rho_0}{d \Lambda} &=& \frac{\Lambda^3}{4 \pi^2}
\left( 3 + 2 \frac{\rho_0 V_0^{(3)}}{V_0''} \right)
\frac{1}{\sqrt{\Lambda^2 + 2 \rho_0 V_0''}} 
\frac{1}{\exp \sqrt{\frac{\Lambda^2}{T^2} + \frac{2 \rho_0 V_0''}{T^2}} -1}
\Theta(\Lambda^2 + 2 \rho_0 V_0'')
\label{feqphi0}
\eea
Furthermore, for a study of the critical properties at a second order
phase-transition, one should make use of the fact that properly rescaled
quantities asymptotically exhibit scaling at the critical temperature. 
We introduce the three-dimensionally rescaled potential and field through
\bea
\kappa = \frac{\rho}{\Lambda T} \qquad ; \qquad 
u = \frac{V}{\Lambda^3 T} \qquad ; \qquad 
\frac{d^i V}{d \rho^i} = \Lambda^{3-i} T^{1-i} \frac{d^i u}{d \kappa^i} 
\label{scalingvars}
\eea
Using $\lambda = \frac{\Lambda}{T}$, the flow-equations for $u$ and $\kappa_0$
now read
\bea
\lambda \frac{d u}{d \lambda} &=& -3 u + u' \kappa - \frac{1}{2
\pi^2} \ln \left\{ 1 - \exp\left[ -\lambda \sqrt{1 + u' + 2 \kappa u''} \right] \right\}
\Theta(1 + u' + 2 \kappa u'')
\label{fequ}
\eea
and
\bea
\lambda \frac{ d \kappa_0}{ d \lambda} = - \kappa_0 +
\frac{\lambda}{4\pi^2} \left( 3 + 2 \frac{\kappa_0 u_0^{(3)}}{u_0''} \right)
\frac{1}{\sqrt{1+2 \kappa_0 u_0''}} \left[
\exp\left( \lambda \sqrt{1+ 2 \kappa_0 u_0''} \right) - 1 \right]^{-1}
\Theta(1+2 \kappa_0 u_0'')
\label{feqvp0}
\eea
The three-dimensional rescaling in (\ref{scalingvars}) is motivated by the
fact that for $\Lambda^2 + V' + 2 \rho V'' \ll T^2$, i.e. when both the
external scale and the mass-scale of the theory are small compared to the
temperature, the flow-equation for $u$, eq. (\ref{fequ}), reduces to the
flow-equation of the purely three-dimensional theory with a $\Theta$-function
cutoff (up to a field-independent contribution): 
We have
\bea
\lambda \frac{d u}{d \lambda} \!=\! -3 u\! +\! u' \kappa \!-\! \left[ \frac{1}{4
\pi^2} \ln (1 \!+\! u'\! +\! 2 \kappa u'') + \frac{1}{2\pi^2} \ln \lambda \right]
\Theta(1 \!+\! u' \!+\! 2 \kappa u'') +
{\mathcal{O}} \left( \sqrt{\frac{\Lambda^2 \!+\! V'\! +\! 2 \rho V''}{T^2}} \right)
\label{feq3d}
\eea
This is the manifestation of dimensional reduction and the dominance of
classical over quantum fluctuations in the critical domain
\cite{Buchmueller}.\vspace*{0.5cm}

\noindent {\bf{3.}} We now proceed to the results for the critical behavior of
the theory obtained by a study of the system of equations resulting from
(\ref{feqV}) - (\ref{feqvp0}).
As was already pointed out in 
\cite{DAP1},
the phase-transition is found to be second order, consistent with the
expectations from other theoretical studies and experiments.
We will in this letter concentrate on the critical exponents and the equation
of state. 
More detailed results on other aspects of the critical behavior like critical
amplitudes and nonuniversal quantities will be presented elsewhere
\cite{JuergenInPrep}.
As a starting potential at $\Lambda_0 \gg T$ (in practice we use
$\Lambda_0 = \exp(3.5) T \approx 30 T$) one needs the effective potential of
the theory at vanishing temperature.
We use here the tree-level potential 
\bea
V_{\Lambda_0}(\rho) = \frac{g(\Lambda_0)}{2} \left( \rho - \rho_0(\Lambda_0) \right)^2
\label{V0}
\eea
and a small value of $g(\Lambda_0) = 0.1$. 
In principle one could use the one- or two-loop potential, but this would only
change the nonuniversal results.
Below, all dimensionful quantities are given in units of $\sqrt{\rho_0}$ which
we arbitrarily set to $1$.

At the critical temperature, the theory exhibits scaling: The dimensionless
minimum of the potential approaches a nonvanishing constant as $\lambda =
\frac{\Lambda}{T} \rightarrow 0$. 
\begin{figure}
\centering
\epsfig{file=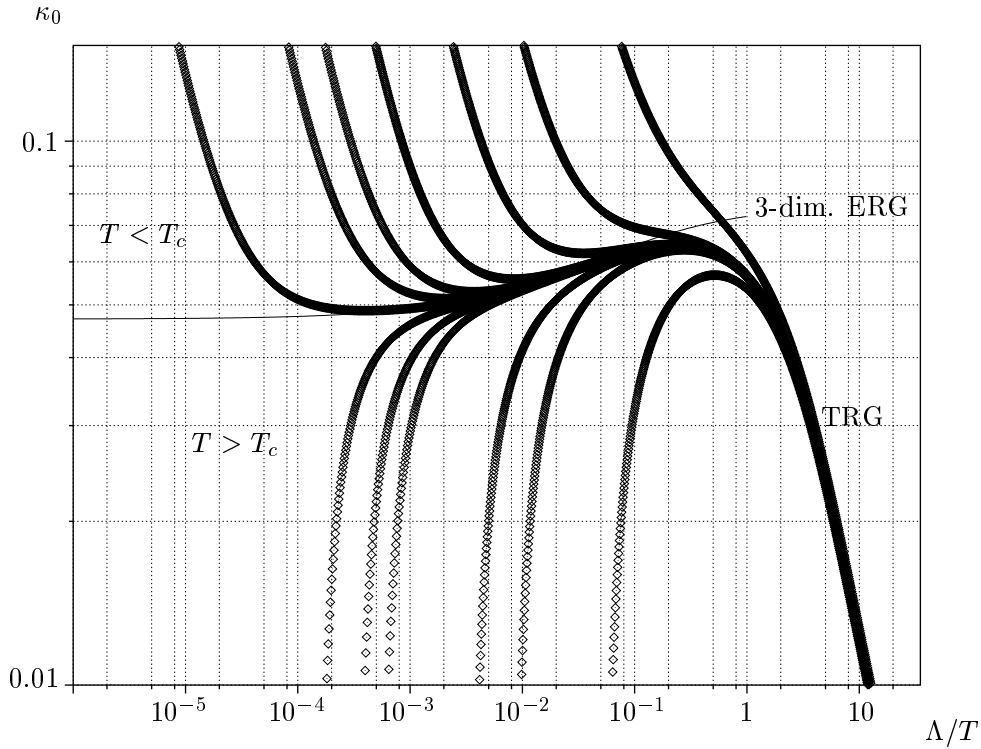,clip=,width=13cm}
\renewcommand{\baselinestretch}{1.2}\normalsize
\caption{The dimensionless minimum of the effective potential as a function of
$\Lambda / T$ for different temperatures above and below the critical one
($g(\Lambda_0) = 0.1$).
The solid line gives the scaling solution of the purely three-dimensional
theory.
Dimensional reduction is nicely observed.}
\renewcommand{\baselinestretch}{1.7}\normalsize
\end{figure}
Figure 1 displays $\kappa_0$ as a function of $\lambda$ for different
temperatures above and below the critical temperature. 
Above the critical
temperature the theory is in the symmetric phase and the minimum reaches $0$
for finite values of $\lambda$. Below the critical temperature, the theory is
in the broken phase and the {\it{dimensionful}} minimum $\rho_0 = \lambda T^2
\kappa_0$ reaches a constant as $\lambda \rightarrow 0$ -- the dimensionless
minimum $\kappa_0$ diverges.
Also included is the result of a study of the effective three-dimensional
theory using the leading contribution of (\ref{feq3d}) only.
In this case, the parameter controlling the deviation from the critical line is
not the temperature but the value of the minimum at the starting scale.
One nicely observes the effective three-dimensional behavior of the
finite-temperature theory close to the phase transition after an initial
running which is governed by the four dimensional theory (in the case of the
TRG the dimensionful parameters are constant in this regime, since the quantum
fluctuations are included in the starting action already) for $\Lambda \gg T$
and a "crossover" regime for $\Lambda \approx T$.
The TRG in this way provides the possibility of studying dimensional reduction
nonperturbatively.

The existence of the asymptotic scaling solution implies universal critical
exponents and a scaling equation of state. 
From the study performed in the present work we can extract all critical
quantities directly in the full finite-temperature theory. 
In the following we concentrate on the exponents $\beta, \gamma$ and $\delta$
and the equation of state.
The approximation performed by neglecting nontrivial terms in the derivative
expansion prevents us from studying the exponent $\eta$ which is given by the
anomalous dimension of the order-parameter field at the critical point. 
We thus by assumption have $\eta = 0$, which has to be compared to the best
value in the literature, giving $\eta = 0.032$
\cite{ZJ}. 
As stated above, the approximation is rather modest in the case under study.
We expect the other critical exponents to have an error of the order of $\eta$,
i.e. of about 5 \%.

\begin{table}
\begin{center}
\begin{tabular}{||c||c|c|c|c|c||}
\hline
 & $\beta$ & $\gamma$ & $\delta$ & $\eta$ & $\nu$ \\
\hline
 This work & 0.35 & 1.32 & 5.0 & 0 & 0.66 $(=\nu/2)$ \\
 TRG+LPA \cite{DAP1} & & & 3.57 & 0.015 & 0.58 \\
\hline
3d ERG + LPA \cite{CN1} & 0.333 & 1.247 & & 0.045 & 0.638 \\
3d ERG \cite{JNC} & 0.336 & 1.258 & 4.75 & 0.044 & 0.643 \\
\hline
Best values \cite{ZJ} & 0.325 & 1.240 & 4.81 & 0.032 & 0.630 \\
\hline
\end{tabular}
\caption{Critical exponents from different approaches}
\end{center}
\end{table}

The exponents we will consider are defined in the following way:
The exponent $\beta$ describes the temperature-dependence of the magnetization
(in our case of the expectation value of $\varphi$) as the temperature
approaches the critical temperature from below: ($\varphi_0 = \sqrt{2 \kappa_0}$)
\bea
\varphi_0 \propto (T_c - T)^\beta
\label{beta}
\eea
The exponent $\gamma$ describes the behavior of the susceptibility, in our
case corresponding to the inverse mass of the order-parameter field, as $T$
reaches $T_c$.
We have
\bea
V'(\rho_0) + 2 \rho_0 V''(\rho_0) \propto | T_c - T |^\gamma
\label{gamma}
\eea
Finally, the exponent $\delta$ describes the form of the effective potential
for small $\rho$ at $T = T_c$ through
\bea
V'(\rho) \propto \rho^{\frac{\delta-1}{2}}
\label{delta}
\eea
Table 1 gives our values for the critical exponents in comparison with results
from other studies. 
There is a clear improvement of the results compared to the exponents found
using the LPA in 
\cite{DAP1}.
Compared to the best known values, the errors are of the expected order.
The results could be further improved by going beyond leading order in the derivative
expansion. 

We finally turn to the equation of state. 
This has been addressed in the three-dimensional theory through the
$\epsilon$-expansion, ERG-calculations, and lattice studies. 
We will compare our results below.
The equation of state is essentially given by the derivative of the effective
potential as a function of the order-parameter field (in a solid-state context
the magnetic field $H$ as a function of the magnetization $M$) in the critical
regime.
\begin{figure}
\centering
\epsfig{file=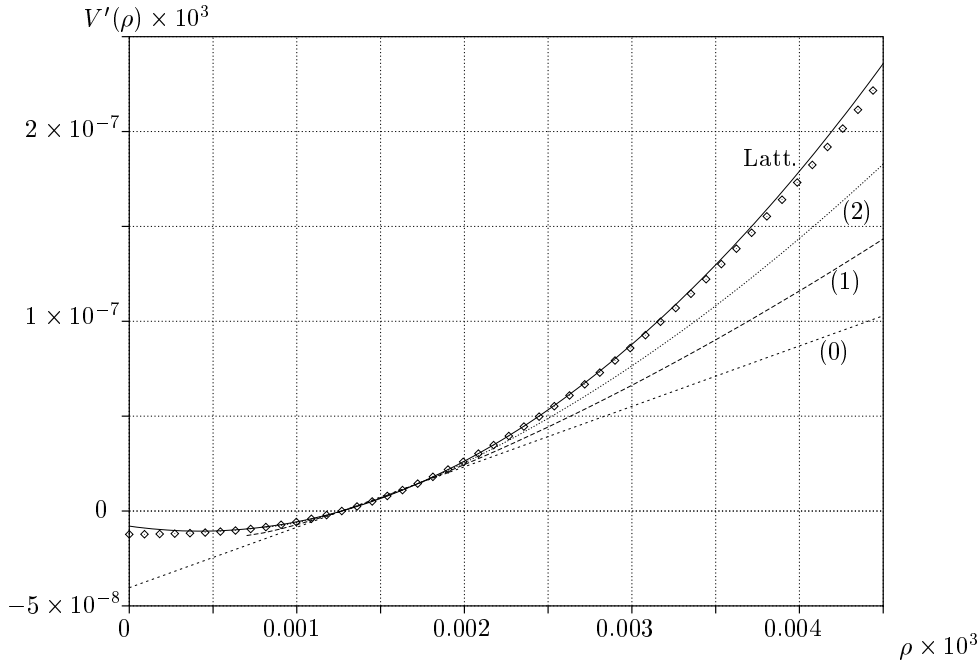,clip=,width=13cm}
\renewcommand{\baselinestretch}{1.2}\normalsize
\caption{$V'(\rho)$ in the critical regime from the thermal renormalization
group (diamonds).
For comparison we give the properly rescaled results of a lattice study 
\cite{Tsypin} and the $\epsilon$-expansion to order $\epsilon^0$, $\epsilon^1$
and $\epsilon^2$
\cite{GuidaZJ}.}
\renewcommand{\baselinestretch}{1.7}\normalsize
\end{figure}
Up to the normalization of the potential and the field, the equation of state
is universal, and may for not too large values of the field be computed in the
three-dimensional theory (for very large field-values, dimensional reduction
generally does not hold since the mass of the order-parameter field is large
compared to the temperature).
We choose here a temperature slightly below the critical one but safe inside
the critical domain and plot $V'(\rho)$ in figure 2 (diamonds).
For comparison, we also display the properly rescaled results of a lattice
study
\cite{Tsypin},
which is up to normalization given by the simple expression
\bea
V'(\rho) = \frac{1}{8} \left( - 1 - 4 \rho  + 12 \rho^2 \right)
\label{vplatt}
\eea
as the solid line as well as the results found from the $\epsilon$-expansion to
order $\epsilon^0$, $\epsilon$ and $\epsilon^2$
\cite{GuidaZJ} (the dashed lines labeled by (0), (1) and (2) resp.).
This figure may be compared to figure 5 in 
\cite{Tsypin},
where also a comparison of the lattice results and the results of the ERG-study
in the three-dimensional theory 
\cite{JNC}
is given. 
One observes an impressive agreement of both the ERG-results of 
\cite{JNC} 
and the TRG-results presented here with the lattice-calculations, while the
$\epsilon$-expansion gives accurate values for the critical exponents but works
only rather modestly for the equation of state.\vspace*{0.5cm}

\noindent{\bf{4.}} In conclusion, we have presented an approximate solution of
the thermal renormalization group-equation for the one-component
selfinteracting scalar theory in the critical domain.
We work in lowest order in the derivative expansion but make no approximations
to the form of the effective potential.
We explicitly obtain a nonperturbative description of dimensional reduction
and demonstrate how the critical behavior of the theory is dominated by
three-dimensional classical fluctuations (figure 1).
The critical exponents corresponding to the second order phase-transition that
the theory exhibits are obtained with an accuracy of the order of 5 \% already
within the relatively simple approximations performed here and could be
improved by going to higher orders in the derivative expansion (table 1).
We finally find results for the equation of state of the theory that are in
surprising agreement with the results of lattice studies (figure 2).

All in all this work demonstrates that the thermal renormalization group is a
valid tool for the study of critical phenomena at finite temperature.
It has a number of advantages compared to other methods, maybe the most
prominent one being the fact that it is manifestly gauge-invariant.
It thus simplifies the study of the critical behavior of gauge theories as
compared to the usual ERG-approach since no non-gauge-invariant counter-terms
are enforced through the introduction of the cutoff.
Furthermore the TRG makes it possible to work with effective actions for
gauge-theories also for nonvanishing values of the external scale,
corresponding to the course-grained free energies used in statistical mechanics
for the study of first order phase transitions. 
Applications to the critical behavior of gauge theories should thus be very
interesting.

\bigskip

\noindent{\bf{Acknowledgment:}} I would like to thank J\"urgen Reingruber,
Chris Stephens and Johannes Manus for interesting discussions.

\end{document}